\documentclass{elsart}

\usepackage{amssymb}
\usepackage{graphics}
\usepackage{version}
\usepackage{color}
\input{epsf}

\begin{document}

\begin{frontmatter}


\title{Beam normal spin asymmetry in elastic lepton-nucleon scattering}

\author[Genova]{M. Gorchtein}
\address[Genova]{Dipartimento di Fisica dell'Universit\`a di Genova, 
Sezione INFN di Genova, I-16146 Genova, Italy}
\author[Saclay]{P.A.M. Guichon}
\address[Saclay]{SPhN/DAPNIA, CEA Saclay, F-91191 Gif sur Yvette, France}
\author[JLab,WM]{M. Vanderhaeghen}
\address[JLab]{Jefferson Laboratory, Newport News, VA 23606, USA}
\address[WM]{Department of Physics, College of William and Mary,
Williamsburg, VA 23187, USA}

\begin{abstract}
We discuss the two-photon exchange contribution to observables 
which involve lepton helicity flip in elastic 
lepton-nucleon scattering.
This contribution is accessed through the  spin asymmetry for a lepton beam 
 polarized normal to the scattering plane. 
We estimate this beam normal spin asymmetry 
at large momentum transfer using a parton model
 and we express the corresponding amplitude in terms 
of generalized parton distributions.
\end{abstract}

\end{frontmatter}

\section{Introduction}
\label{sec:intro}

\begin{figure}
\epsfxsize=16.cm
\centerline{\epsffile{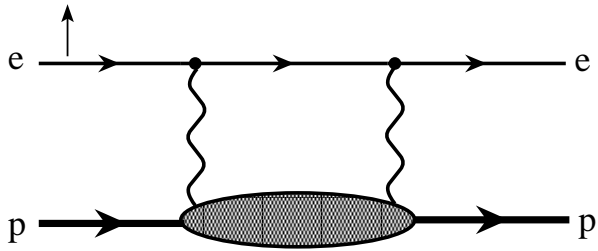}}
\vspace{-7cm}
\caption{Two-photon exchange amplitude entering the elastic 
lepton-nucleon scattering, 
with a beam spin polarized normal to the scattering plane.} 
\label{fig:twophoton}
\end{figure}

Elastic electron-nucleon scattering in the one-photon exchange approximation
gives  direct access to the electromagnetic  form factors of the nucleon, 
 an essential piece of information about its structure. 
In recent years, the ratio $G_{E p} / G_{M p}$ 
of the proton's electric to 
magnetic form factors has been measured up to large momentum 
transfer $Q^2$ in precision experiments~\cite{Jones00,Gayou02} using 
the polarisation transfer method.
 It came as a surprise that 
these experiments for $Q^2$ up to 5.6 GeV$^2$
extracted a ratio of $G_{E p} / G_{M p}$ which is incompatible with 
unpolarized measurements \cite{Slac94,Chr04,Arr03} 
using the Rosenbluth separation technique. 
\newline
\indent
It has been suggested on general grounds 
in Ref.~\cite{GV03} that this puzzle may be 
explained by a two-photon exchange amplitude (see Fig.~\ref{fig:twophoton}) 
whose magnitude is a few percent 
of the one photon exchange term. 
The failure of the one-photon approximation for  
elastic electron-nucleon scattering is amplified at large $Q^2$ in the 
case of the Rosenbluth extraction of $G_{E p} / G_{M p}$ , but affects 
the polarization method only little. 
A model calculation of the two-photon exchange amplitude when the 
hadronic intermediate state is a nucleon was performed 
in Ref.~\cite{BMT03}. It found that the $2 \gamma$ exchange correction with 
intermediate nucleon can partially resolve the discrepancy between the two 
experimental techniques. 
Very recently, the two-photon exchange contribution to elastic 
electron-nucleon scattering has been estimated 
at large momentum transfer~\cite{YCC04}, 
through the scattering off a parton in the proton by relating 
the process on the nucleon to the generalized parton distributions. 
This calculation found that the $2 \gamma$ exchange contribution is 
indeed able to quantitatively resolve the discrepancy between 
Rosenbluth and polarization transfer experiments. 
\newline
\indent 
To use electron scattering as a precision tool, one needs a  good control
 of the two-photon exchange mechanisms, 
which justifies a systematic study of these  effects, both
 theoretically and experimentally.
The real (dispersive) part of the two-photon exchange amplitude 
can be accessed through the difference between 
elastic electron and positron scattering off a nucleon. 
The imaginary (absorptive) part of the two-photon exchange amplitude 
on the other hand can be accessed through a single spin asymmetry (SSA) in 
elastic electron-nucleon scattering, when either the target or beam spin 
are polarized normal to the scattering plane. 
As time reversal invariance forces 
this SSA to vanish in the   one-photon exchange  approximation, 
it requires the exchange of at least two photons between lepton and nucleon 
(see Fig.~\ref{fig:twophoton}), and 
hence it is of order $\alpha = e^2 / (4 \pi) \simeq 1/ 137$. 
\newline
\indent
In the context of QED, such normal spin asymmetries have been 
discussed long time ago~\cite{BF60}. In Ref.~\cite{BF60}, 
the leading order (box diagram) calculations for 
$\e^{- \uparrow} \mu^-$ and $e^{- \uparrow} e^-$ (M\o{}ller) 
elastic scattering with an electron polarized normal to the 
scattering plane have been performed. 
The resulting SSA is directly sensitive to a QED rescattering phase.  
Recently, the E158 experiment at SLAC ~\cite{E158} has measured 
the M\o{}ller scattering $e^{- \uparrow} e^- \to e^- e^-$ 
with normal beam spin polarization, allowing for a precsion 
comparison with the latest QED calculations ~\cite{DS04}, including 
leading logarithmic QED corrections due to intial and final state 
radiation effects.  
\newline
\indent
To get an order of magnitude estimate for the beam normal SSA, we have 
to consider that, besides the reduction by an overall factor $\alpha$, the    
polarization of an ultra-relativistic electron in the direction 
normal to its momentum involves 
a suppression factor $m_e / E_e$ (with $E_e$ the electron beam energy), 
which is of order $10^{-4}$ to $10^{-3}$ for beam energies 
in the few GeV range. 
Therefore, the beam normal SSA is of order $10^{-6}$ 
(for some order of magnitude estimates, see Refs.~\cite{RKR71,AAM02}). 
A measurement of such small asymmetries is quite demanding experimentally. 
However, in the case of a polarized lepton beam, asymmetries of the order ppm 
are currently accessible in parity violation (PV) elastic electron-nucleon 
scattering experiments. 
The parity violating asymmetry involves a beam spin aligned 
in the direction of its momentum. The SSA for an electron 
beam spin normal to the scattering plane, which corresponds 
with a  flip of the lepton helicity, 
can also be measured using the same experimental set-ups. 
First measurements of this beam normal SSA at beam energies below 1 GeV 
have yielded values as large 
as 10 ppm in magnitude~\cite{Wells01,Maas03,E158}. 
At higher beam energies, the beam normal SSA can be measured in upcoming  
PV elastic electron-nucleon scattering experiments~\cite{Happex,G0}. 
One may therefore envisage the possibility in near future to access this 
asymmetry at larger momentum transfer ( for $Q^2$ around or larger 
than 1~GeV$^2$), where one may expect that the scattering off a parton starts 
to dominate. 
\newline
\indent
The aim of the present work is to develop the formalism for elastic 
electron-nucleon scattering  with a  flip of the lepton helicity. 
To provide an estimate for the two-photon amplitudes, 
we extend the partonic calculation 
of Ref.~\cite{YCC04}. 
In this partonic calculation, the real and imaginary parts of the 
$2 \gamma$ exchange amplitudes are related. Hence in the partonic
 regime, 
a direct comparison of the imaginary part with experiment 
can also provide a very valuable cross-check 
on the calculated result for the real part.  
\newline
\indent
In Section~\ref{sec:elasticen}, we develop the formalism for elastic 
lepton-nucleon scattering which involves a  flip of the lepton helicity.  
In Section~\ref{sec:quark}, we calculate the two-photon exchange
lepton-quark scattering amplitude.
We then construct in Section~\ref{sec:nucleon} 
the lepton helicity flip amplitudes for the nucleon 
through a convolution of the quark amplitudes with a generalized 
parton distribution. 
We discuss our results in Section~\ref{sec:results} and conclude in 
Section~\ref{sec:concl}.

\section{Elastic lepton-nucleon scattering 
formalism beyond one-photon exchange}
\label{sec:elasticen}

In this work, we consider the elastic lepton-nucleon
scattering~:
\begin{eqnarray}
\label{Eq:intro.2}
l(k)+N(p)\rightarrow l(k')+N(p'),
\end{eqnarray}
for which we adopt the definitions~:
\begin{eqnarray}
\label{Eq:intro.3}
P=\frac{p+p'}{2},\, K=\frac{k+k'}{2},\, q=k-k'=p'-p,
\end{eqnarray}
and choose 
$Q^{2}=-q^{2}$ and $\nu =K.P$ 
as the independent invariants of the scattering. 
They are related to the Mandelstam invariants $s = (k + p)^2$ and 
$u = (k' - p)^2$ through~: $s - u = 4 \nu$ and $s + u = Q^2 + 2 M^2$, 
where $M$ is the nucleon mass. 
For further use, we also introduce the usual polarization 
parameter $\varepsilon$ of the virtual photon, which   
can be related to the invariants $\nu$ and $Q^2$ 
as (neglecting the lepton mass $m_l$)~:
\begin{eqnarray}
\label{Eq:intro.6}
\varepsilon =\frac{\nu ^{2}-M^{4}\tau (1+\tau )}{\nu ^{2}+M^{4}\tau (1+\tau )},
\end{eqnarray}
with $\tau = Q^2 / (4 M^2)$. 
\indent
For a theory which respects Lorentz, parity and charge
conjugation invariance, the general amplitude for elastic scattering
of two spin 1/2 particles depends on six invariant amplitudes as given
by Goldberger {\it et al.}~\cite{Goldb57}. The amplitude can be
expanded in the following set of invariants~:
\begin{eqnarray}
\bar{u}(k') u(k) &\; \; \; \cdot \; \; \;& \bar{u}(p') u(p), 
\nonumber\\
\bar{u}(k') u(k) &\; \; \; \cdot \; \; \;& \bar{u}(p') \gamma . K u(p), 
\nonumber\\
\bar{u}(k') \gamma_5 u(k) &\; \; \; \cdot \; \; \;& \bar{u}(p') \gamma_5 u(p), 
\nonumber\\
\bar{u}(k') \gamma . P u(k) &\; \; \; \cdot \; \; \;& 
\bar{u}(p') \gamma . K u(p), \nonumber\\
\bar{u}(k') \gamma . P u(k) &\; \; \; \cdot \; \; \;& 
\bar{u}(p') u(p), \nonumber\\
\bar{u}(k') \gamma_5 \gamma . P u(k) &\; \; \; \cdot \; \; \;& 
\bar{u}(p') \gamma_5 \gamma . K u(p),
\label{eq:gold}
\end{eqnarray}
where the last three structures conserve the helicity of the lepton 
and the first three flip it (i.e. are of the
order of the mass of the lepton, $m_l$).
Using the Dirac equation and elementary relations among
the Dirac matrices, the last structure in Eq.~(\ref{eq:gold}) can be
traded for 
$\bar{u}(k') \gamma_{\mu} u(k) \cdot \bar{u}(p') \gamma^{\mu} u(p)$, 
and a combination of the first five structures. 
Therefore, one can write the general elastic lepton-nucleon scattering
amplitude as~:
\begin{eqnarray}
T \;=\; T^{non-flip} \;+\; T^{flip} ,
\label{eq:tampl}
\end{eqnarray}
with \cite{GV03}~:
\begin{eqnarray}
\label{eq:non-flip}
T_{h' \lambda'_N, \, h \lambda_N}^{non-flip} \,&=&\, 
\frac{e^{2}}{Q^{2}} \, \bar{u}(k', h')\gamma _{\mu }u(k, h)\, 
\nonumber \\
&\times &\,
\bar{u}(p', \lambda'_N)\left( \tilde{G}_{M}\, \gamma ^{\mu }
-\tilde{F}_{2}\frac{P^{\mu }}{M}
+\tilde{F}_{3}\frac{\gamma .KP^{\mu }}{M^{2}}\right) u(p, \lambda_N),
\end{eqnarray}
and~:
\begin{eqnarray}
\label{eq:flip}
\hspace{-0.4cm}
T_{h' \lambda'_N, \, h \lambda_N}^{flip} \,&=&\, 
\frac{e^{2}}{Q^{2}} \frac{m_l}{M}
\left[ \, \bar{u}(k', h') u(k, h)\, \cdot \,
\bar{u}(p', \lambda'_N)\left( \tilde{F}_{4}\, 
+\tilde{F}_{5}\frac{\gamma . K}{M} \right) u(p, \lambda_N) \right. \nonumber \\
&&\hspace{1.25cm} \left. + \, 
\tilde{F}_{6}\,\bar{u}(k', h') \gamma_5 u(k, h)\, \cdot \,
\bar{u}(p', \lambda'_N) \gamma_5 u(p, \lambda_N) \right] ,
\end{eqnarray}
where $h (h')$ are the helicities of the incoming (outgoing) leptons, 
$\lambda_N (\lambda'_N)$ 
are the helicities of the incoming (outgoing) nucleons,  
and where $m_l$ denotes the lepton mass. 
In Eqs.~(\ref{eq:non-flip}) and (\ref{eq:flip}), 
$\tilde{G}_{M}$, $\tilde{F}_{2}$, $\tilde{F}_{3}$,
$\tilde{F}_{4}$, $\tilde{F}_{5}$, $\tilde{F}_{6}$ are 
complex functions of $ \nu $ and $ Q^{2} $, and 
the factor $ e^{2}/Q^{2} $ has been introduced for convenience, 
with $e$ the proton charge.  
Furthermore in Eq.~(\ref{eq:flip}), we extracted an explicit factor 
$m_l / M$ out of the amplitudes, which reflects the fact that 
for a vector interaction (such as in QED), 
the lepton helicity flip amplitude 
vanishes when $m_l \to 0$. 
In the Born approximation, one obtains~:
\begin{eqnarray}
\label{eq:born}
\tilde{G}_{M}^{Born}(\nu ,Q^{2}) \,&=&\, G_{M}(Q^{2}),   \nonumber\\
\tilde{F}_{2}^{Born}(\nu ,Q^{2}) \,&=&\, F_{2}(Q^{2}),   \nonumber\\
\tilde{F}_{3, \, 4, \, 5, \, 6}^{Born}(\nu ,Q^{2}) \,&=&\, 0,  	
\end{eqnarray}
where $G_M(Q^2)$ and $F_2(Q^2)$ are the magnetic and Pauli form factors 
respectively. 
Since $\tilde{F}_{3}$, $\tilde{F}_{4}$, $\tilde{F}_{5}$, $\tilde{F}_{6}$,  
and the phases of \( \tilde{G}_{M} \) and \( \tilde{F}_{2} \) 
vanish in the Born approximation, they must
originate from processes involving at least the exchange of two photons.
Relative to the factor \( e^{2} \) introduced in 
Eq.~(\ref{eq:non-flip}), we see that these are at least of order \( e^{2}. \)
For convenience, we trade the invariant $\tilde F_2$ for $\tilde G_E$,
defined as~:
\begin{eqnarray}
\label{Eq:Obs.11}
\tilde{G}_{E} \equiv \tilde{G}_{M}-(1+\tau )\tilde{F}_{2},
\end{eqnarray}
In the Born approximation $\tilde G_E$ reduces to the
electric form factor $G_E(Q^2)$.
\newline
\indent
For a beam polarized normal to the scattering plane, we can
define a single spin asymmetry, 
\begin{eqnarray}
B_n \,=\, 
\frac{\sigma_\uparrow-\sigma_\downarrow}{\sigma_\uparrow+\sigma_\downarrow}\,,
\label{eq:tasymm}
\end{eqnarray} 
where $\sigma_\uparrow$ ($\sigma_\downarrow$) denotes the cross section 
for an unpolarized target and for a lepton beam spin 
parallel (anti-parallel) to the normal polarization vector, defined
as~:
\begin{eqnarray}
S_n^\mu \,=\, (\,0\,,\, \vec S_n \,), \hspace{2cm}
\vec S_n \,\equiv \,  (\vec{k}\times\vec{k}') \,/\, | \vec{k}\times\vec{k}' | .
\label{eq:sn}
\end{eqnarray}
We refer to this asymmetry as the beam normal
spin asymmetry ($B_n$). 
Its leading non-vanishing contribution is linear in the lepton mass. 
Furthermore, $B_n$ vanishes in the Born approximation, and is therefore 
of order $e^2$. 
Keeping only the leading term of order $e^2$, $B_n$ arises 
from an interference between the one- and two-photon exchange amplitudes.  
In terms of the invariants of Eqs.~(\ref{eq:non-flip},\ref{eq:flip}), 
$B_n$ is given by~:
\begin{eqnarray}
B_n \,&=&\, \frac{2 \, m_l}{Q} \, 
\sqrt{2 \, \varepsilon \, (1-\varepsilon )} \, \sqrt{1 + \frac{1}{\tau}} \, 
\left( G_M^2 \,+\, \frac{\varepsilon}{\tau} \, G_E^2 \right)^{-1} \nonumber\\
&\times& \left\{ - \tau \, G_M \, 
{\mathcal I}\left( \tilde F_3 
+ \frac{1}{1 + \tau} \,\frac{\nu}{M^2} \, \tilde F_5 \right)  
\, - \, G_E \, {\mathcal I} \left( \tilde{F}_{4}  
+ \frac{1}{1 + \tau} \, \frac{\nu}{M^2} \, \tilde F_5 \right) 
\right\}  \nonumber \\
\,&+& \, {\mathcal{O}}(e^4) , 
\label{eq:bnsa2} 
\end{eqnarray}
where ${\mathcal I}$ denotes the imaginary part.

\section{Imaginary part of elastic lepton-quark scattering}
\label{sec:quark}

To evaluate the two photon exchange amplitudes, we decompose the lower blob
 in Fig.~\ref{fig:twophoton} into a part where the intermediate state 
is a nucleon, which we call the nucleon pole part, 
and the rest which we call the inelastic part. The nucleon pole part 
is exactly calculable since it involves only the on shell form factors 
and we discuss only the inelastic part 
which we model by the handbag diagram (see Fig.~\ref{fig:handbag}) 
where the photon scatters off a quark which is 
approximately on the mass shell. The first step is the evaluation of the 
elastic lepton-quark scattering~:
\begin{equation}
\label{eq:eqscatt}
l(k)+q(p_q)\rightarrow l(k')+q(p'_q),
\end{equation}
which involves two independent kinematical invariants~:
$\hat s \equiv (k + p_q)^2$ and $Q^2 = -(k - k')^2$. For further
use, we also introduce the crossing variable 
$\hat u \equiv (k - p'_q)^2$, which satisfies $\hat s + \hat u = Q^2$ 
(for massless quarks).
The $T$-matrix for the lepton-quark hard scattering process 
can be written as~:
\begin{eqnarray}
\label{eq:tmatrixhard}
H^{hard}_{h' h, \, \lambda} &=& 
\frac{(e \, e_q)^2}{Q^{2}} \left\{ 
\bar{u}(k', h')\gamma _{\mu }u(k, h) \cdot    
\bar{u} (p'_q, \lambda) \left( \, \tilde{f}_{1} \, \gamma ^{\mu }
+ \tilde{f}_{3} \, \gamma .K \, P_q^{\mu } \, \right) u(p_q, \lambda)
\right. \nonumber \\
&& \hspace{1cm}+ \left. m_l \,\,\tilde{f}_5 \,\, \bar{u}(k', h') u(k, h)
\cdot  \bar{u} (p'_q, \lambda) \gamma . K u(p_q, \lambda) \right\} ,
\end{eqnarray}
with $P_q \equiv (p_q + p'_q) / 2$, 
where $e_q$ is the charge of the quark in units of $e$,
and where $u(p_q, \lambda)$ and $u(p'_q, \lambda)$ are the quark
spinors with quark helicity $\lambda = \pm 1/2$, which is
conserved in the hard scattering process. In Eq.~(\ref{eq:tmatrixhard}),  
$\tilde{f}_{1}, \tilde{f}_3$ and $\tilde{f}_{5}$ are 
the invariant amplitudes for the scattering of leptons off massless
quarks ($m_q = 0$). They are the analogues of the invariants introduced in 
Eqs.~(\ref{eq:non-flip}, \ref{eq:flip}) 
for the nucleon except that, for obvious reasons, 
we do not introduce powers of the quark mass to make them dimensionless. 
In the one-photon exchange approximation $\tilde{f}_1 \to 1$, and 
$\tilde{f}_3 \to 0, \tilde{f}_5 \to 0$.
Note that for massless quarks, 
quark helicity conservation leads to the 
absence of analogues of  $\tilde F_2, \tilde F_4$ and $\tilde F_6$.
\newline
\indent
To calculate the beam normal spin asymmetry $B_n$, 
we need the corresponding expressions for the 
imaginary parts of $\tilde f_1, \tilde f_3$ and $\tilde f_5$. 
These imaginary parts 
originate solely from the direct two-photon exchange box diagram 
on the quark level.  
The amplitudes $\tilde f_1$ and $\tilde f_3$,  
which conserve the lepton helicity, were 
already calculated in Ref.~\cite{YCC04}.
It was shown in that work that the amplitude $\tilde f_1$ can be separated 
into a soft and hard part, i.e. 
$\tilde{f}_1 = \tilde{f}_1^{soft} + \tilde{f}_1^{hard}$. 
The soft part corresponds with the situation where 
one of the photons in Fig.~\ref{fig:twophoton} carries zero four-momentum, 
and is obtained by replacing the other photon's four-momentum by 
$q$ in both numerator and denominator of the loop integral. 
This yields (for $\hat s > 0$ and $\hat u < 0$) the expressions~\cite{YCC04}~:
\begin{eqnarray}
{\mathcal I}\left( \tilde{f}_1^{soft} \right)
&=& - \frac{e^2}{4 \pi} \, 
\ln \left( \frac{\lambda^2}{\hat s} \right),  
\label{eq:f1qimsoft} \\
{\mathcal I}\left( \tilde{f}_1^{hard} \right)
&=& - \frac{e^2}{4 \pi} 
\left\{ \frac{Q^2}{2 \, \hat u}  
\ln \left( \frac{\hat s}{Q^2}  \right) 
+ \frac{1}{2} \right\}, 
\label{eq:f1qimhard} \\
{\mathcal I}\left( \tilde{f}_3 \right)
&=& - \frac{e^2}{4 \pi} 
\, \frac{1}{\hat u} \, \left\{ 
\frac{\hat s - \hat u}{\hat u}  
\ln \left( \frac{\hat s}{Q^2}  \right) 
+ 1 \right\}.  
\label{eq:f3qim}
\end{eqnarray}
Note that the IR divergent part in Eq.~(\ref{eq:f1qimsoft}), 
proportional to the fictitious photon mass $\lambda^2$,  
does not contribute when calculating physical observables as it just 
corresponds to the lowest order calculation of the Coulomb phase of 
the amplitude. 
\newline
\indent
Analogously to Eqs.~(\ref{eq:f1qimsoft}, \ref{eq:f1qimhard}, \ref{eq:f3qim}), 
we can calculate the imaginary part of the amplitude 
$\tilde f_5$ and obtain 
(for $\hat s > 0$ and $\hat u < 0$)~:
\begin{eqnarray}
{\mathcal I}\left( \tilde{f}_5 \right)_{2 \gamma}
&=& - \frac{e^2}{4 \pi} 
\, \frac{Q^2}{\hat u} \, 
\left\{ - \frac{1}{\hat u} \ln \left( \frac{\hat s}{Q^2}  \right) 
- \frac{1}{\hat s} \right\}.   
\label{eq:f5qim}
\end{eqnarray}
\newline
\indent
As a useful check we use the previous results to evaluate 
the beam normal spin asymmetry $B_n$ on a quark.  
In the limit of massless quarks, $B_n$ is given (for $e_q = +1$) by~:
\begin{eqnarray}
B_n = \frac{m_l}{Q} \, \sqrt{2 \varepsilon (1 - \varepsilon)} \frac{1}{2}
\, \left\{ - Q^2 \, {\mathcal I}  \left(\tilde{f}_3 \right)_{2 \gamma}
- (\hat s - \hat u) \, {\mathcal I} \left( \tilde{f}_5 \right)_{2 \gamma} 
\right\} .   
\label{eq:bnq1} 
\end{eqnarray}
Using the lepton-quark amplitudes of Eqs.~(\ref{eq:f3qim},\ref{eq:f5qim}), 
this yields~:
\begin{eqnarray}
B_n \,&=&\, \frac{e^2}{4 \pi} \,\, \frac{m_l}{Q} \,\, 
\sqrt{2 \varepsilon (1 - \varepsilon)} \,\,  
\frac{Q^2}{2 \, \hat s} 
\,=\, \frac{e^2}{4 \pi} \,\, \frac{m_l}{Q} \,\, 
\sqrt{\frac{- \hat u}{\hat s}} \,\, \frac{Q^4}{\hat s^2 + \hat u^2}, 
\label{eq:bnq2} 
\end{eqnarray}
where the last step has been obtained by using 
$\varepsilon = - 2 \hat s \hat u \,/\, (\hat s^2 + \hat u^2)$.
The expression of Eq.~(\ref{eq:bnq2}) agrees with the well known 
result for the beam normal SSA for $e^{- \uparrow} \mu^- \to e^- \mu^-$ 
derived long time ago in Ref.~\cite{BF60} (see Eq.~(15) in that work, by 
taking the muon mass equal to zero). 

\section{Imaginary part of elastic lepton-nucleon scattering 
in terms of generalized parton distributions}
\label{sec:nucleon}

\begin{figure}
\epsfxsize=15.cm
\centerline{\epsffile{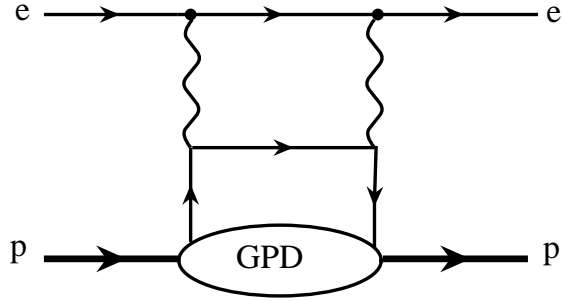}}
\vspace{-5cm}
\caption{Handbag contribution to the two-photon exchange 
amplitude entering the normal beam asymmetry for elastic 
lepton-nucleon scattering.  
The lower blob represents the GPDs of the nucleon.}
\label{fig:handbag}
\end{figure}

Having discussed the two-photon exchange amplitude on the quark, 
we calculate the corresponding amplitudes on the nucleon, as 
is shown in Fig.~\ref{fig:handbag}.
We follow Ref.~\cite{YCC04} and calculate the amplitudes  
in the kinematical regime where  
$s, -u$ and $Q^2$ are large compared to a hadronic scale
($s, -u, Q^2 >> M^2$) as a convolution between a hard scattering 
electron-quark amplitude and a soft matrix element on the nucleon. 
For this kinematical regime, it is convenient to choose 
a frame where $q^+ = 0$, as in~\cite{DY70}, 
where we introduce light-cone variables 
$a^\pm = (a^0 \pm a^3) / \sqrt{2}$ and choose the $z$-axis along the
direction of $P^3$ (so that $P$ has a large + component). 
The lepton + momentum fractions are given by 
$\eta = K^+ / P^+$, with 
\begin{eqnarray}
\eta = 
\left( s - u - 2 \, \sqrt{M^4 - s \, u} \, \right) / (Q^2 + 4 \, M^2).
\nonumber
\end{eqnarray}
In the frame $q^+ = 0$, the
parton light-cone momentum fractions are then defined as  
$x = p_q^+ / P^+ = p_q^{' \, +} / P^+$.   
The active partons, on which the hard scattering takes place,
are approximately on-shell, and their  
intrinsic transverse momenta (defined in a frame where the hadron has
zero transverse momentum) are small and can be neglected
when evaluating the hard scattering process. 
The Mandelstam variables for the process (\ref{eq:eqscatt}) on the quark, 
which enter in the evaluation of the hard scattering amplitude, 
are given by~: 
\begin{eqnarray} 
\label{eq:mandq}
\hat s = \frac{(x + \eta)^2}{4 \, x \, \eta} \, Q^2 \, , \hspace{1cm} 
\hat u = - \frac{(x - \eta)^2}{4 \, x \, \eta} \, Q^2 .
\end{eqnarray}  
The helicity amplitudes for elastic electron-nucleon scattering 
in the kinematical regime where 
$s, -u, Q^2 >> M^2$, can be expressed as~\cite{YCC04}~:
\begin{eqnarray}
\label{eq:handbag}
T_{h' \lambda'_N, \, h \lambda_N}^{hard} &=&
 \int_{-1}^1 \frac{dx}{x} \, \sum_q  \, \frac{1}{2} 
\left[ H_{h' h, + \frac{1}{2}}^{hard} + H_{h' h, - \frac{1}{2}}^{hard} \right] 
\nonumber \\
&& \times \left\{ \left[\, H^q\left(x, 0, q^2 \right) 
\,+\, E^q \left(x, 0, q^2 \right) \right] \, 
\, \frac{1}{2} \, 
\bar{u}(p', \lambda'_N) \, \gamma \cdot n \, u(p, \lambda_N) \right. 
\nonumber \\
&&\left. -\, E^q\left(x, 0, q^2 \right)  \, 
\, \frac{1}{2 M} \, 
\bar{u}(p', \lambda'_N) \, u(p, \lambda_N) \,\right] \nonumber \\
&+&  \int_{-1}^1 \frac{dx}{x} \, \sum_q  \, \frac{1}{2}
\left[ H_{h' h, + \frac{1}{2}}^{hard} - H_{h' h, - \frac{1}{2}}^{hard} 
\right] \nonumber\\ 
&& \times \, \mathrm{sgn}(x) \,
\tilde H^q\left(x, 0, q^2 \right) \frac{1}{2} \,  
\bar{u}(p', \lambda'_N) \, \gamma \cdot n \, \gamma_5 \, u(p, \lambda_N),
\end{eqnarray}
where $H^{hard}_{h' h, \lambda}$ are the hard scattering amplitudes, 
and $n^\mu$ is a Sudakov four-vector ($n^2 = 0$), which is given by~:
\begin{eqnarray}
n^\mu \,=\, 2 /(\sqrt{M^4 -  s u}) \, 
\left\{- \eta \, P^\mu \,+\, K^\mu \right\}.
\end{eqnarray}
Furthermore in Eq.~(\ref{eq:handbag}),  
$H^q, E^q, \tilde H^q$ are the 
Generalized Parton Distributions (GPDs) for a quark $q$ in the
nucleon (for a review, see e.g. Ref.~\cite{GPV01}).
The hard scattering amplitudes $H^{hard}_{h' h, \lambda}$ 
in Eq.~(\ref{eq:handbag}) can be expressed  
in terms of $\tilde f_1^{hard}, \tilde f_3, \tilde f_5$ 
using Eq.~(\ref{eq:tmatrixhard}) as~:
\begin{eqnarray}
\frac{1}{2} \, 
\left[ H_{h' h, + \frac{1}{2}}^{hard} + H_{h' h, - \frac{1}{2}}^{hard} \right] 
&=& \frac{(e \, e_q)^2}{Q^2} \left\{\, \delta_{h', h} \, 
\left[ \tilde f_1^{hard} \, (\hat s - \hat u) - \tilde f_3 \, 
\hat s \hat u \right] \right. \nonumber \\
&& \hspace{-1cm}-\left. \,  \delta_{h', -h} \, (2 h) \, m_l \, \sqrt{Q^2} 
\sqrt{- \hat s \hat u}
\left[ \frac{2}{\hat s} \tilde f_1^{hard} \,+\, \tilde f_3 
\,+\, \tilde f_5 \right] \right\} ,
\label{eq:helavhard} \\
\frac{1}{2} \, 
\left[ H_{h' h, + \frac{1}{2}}^{hard} - H_{h' h, - \frac{1}{2}}^{hard} \right] 
&=& \frac{(e \, e_q)^2}{Q^2} \, \delta_{h', h} \, (2 h) \, 
\tilde f_1^{hard} \, Q^2,
\label{eq:heldiffhard}
\end{eqnarray}
with $\hat s$ and $\hat u$ according to Eq.~(\ref{eq:mandq}).
\newline
\indent
To extract $\tilde F_3, \tilde F_4, \tilde F_5$, we first need to express them 
in terms of the electron-nucleon helicity amplitudes, which is done in 
Appendix \ref{eq:app1}. 
Using Eq.~(\ref{eq:handbag}) and Eqs.~(\ref{eq:f3thel})-(\ref{eq:f5thel}),  
we can finally express $\tilde F_3, \tilde F_4, \tilde F_5$ 
(after some algebra) in terms of the nucleon GPDs as~:
\begin{eqnarray}
\tilde F_3 &=& \frac{M^2}{\nu}
\left( \frac{1 + \varepsilon}{2 \varepsilon} \right) \, (A - C),
\label{eq:F3handbag} \\
\tilde F_4 &=& 
\frac{1}{1 + \tau} \left\{ \sqrt{\frac{1 + \varepsilon}{2 \, \varepsilon}} \, 
\left[ \frac{s + M^2}{s - M^2} \sqrt{\frac{1 + \varepsilon}{2 \varepsilon}} 
A - A'  \right] 
\,-\,\left[\frac{s + M^2}{s - M^2} \sqrt{\frac{1 + \varepsilon}{2 \varepsilon}}
 B - B'  \right] 
\right\} \nonumber \\
&-& \frac{M^2}{\nu} \left( \frac{1 + \varepsilon}{2 \varepsilon} \right) \, C,
\label{eq:F4handbag} \\
\tilde F_5 &=& 
- \frac{M^2}{\nu} \sqrt{\frac{1 + \varepsilon}{2 \, \varepsilon}} \, 
\left[ \frac{s + M^2}{s - M^2} \sqrt{\frac{1 + \varepsilon}{2 \varepsilon}} 
A - A'  \right] 
\nonumber \\
&+& \left(\frac{M^2}{\nu}\right)^2 
\left( \frac{1 + \varepsilon}{2 \varepsilon} \right) \, (1 + \tau) \, C,
\label{eq:F5handbag} 
\end{eqnarray}
where we introduced the integrals containing the GPDs~:
\begin{eqnarray}
A \, &\equiv& \, \int_{-1}^1 \frac{dx}{x}     
\frac{\left[(\hat s - \hat u) \tilde{f}_1^{hard} -
\hat s \hat u \tilde{f}_3 \right]}{(s - u)} \, 
\sum_q e_q^2 \, \left( H^q + E^q \right), 
\nonumber \\
B \, &\equiv& \, \int_{-1}^1 \frac{dx}{x}     
\frac{\left[(\hat s - \hat u) \tilde{f}_1^{hard} - 
\hat s \hat u \tilde{f}_3 \right]}{(s - u)} \, 
\sum_q e_q^2 \, \left( H^q - \tau E^q \right), 
\nonumber \\
C \, &\equiv& \, \int_{-1}^1 \frac{dx}{x} \, \tilde{f}_1^{hard} \, 
\mathrm{sgn}(x) \, \sum_q e_q^2 \, \tilde H^q, 
\nonumber \\
A' \, &\equiv& \, \int_{-1}^1 \frac{dx}{x} \frac{\sqrt{- \hat s \hat u}}{2}    
\left[\frac{2}{\hat s} \, \tilde f_1^{hard} + \tilde f_3 + \tilde f_5 \right] 
\, \sum_q e_q^2 \, \left( H^q + E^q \right), 
\nonumber \\
B' \, &\equiv& \, \int_{-1}^1 \frac{dx}{x} \frac{\sqrt{- \hat s \hat u}}{2}    
\left[\frac{2}{\hat s} \, \tilde f_1^{hard} + \tilde f_3 + \tilde f_5 \right] 
\, \sum_q e_q^2 \, \left( H^q - \tau E^q \right) . 
\label{eq:integrals}
\end{eqnarray}
The expression for $\tilde F_3$ and the quantities $A, B, C$
have been given previously in Ref.~\cite{YCC04}. 
Eqs.~(\ref{eq:F3handbag}-\ref{eq:F5handbag}) reduce to 
the partonic amplitudes in the limit $M \to 0$ by considering a quark  
target, for which the GPDs are given by : 
$H^q \to \delta(1 - x)$, $E^q \to 0$, and $\tilde H^q \to \delta(1 - x)$. 
In this limit, and using the identity $-\hat s \hat u = 4 \nu^2 
\left(\frac{2  \varepsilon}{1 + \varepsilon} \right)$, 
we then easily recover that $\tilde F_3 / M^2 \to e_q^2 \, \tilde f_3$, 
$\tilde F_4 / M \to 0$, and $\tilde F_5 / M^2 \to e_q^2 \, \tilde f_5$. 
\newline
\indent
Inserting the above expressions of 
Eqs.~(\ref{eq:F3handbag}-\ref{eq:F5handbag}) 
for $\tilde F_3$, $\tilde F_4$ and $\tilde F_5$ into Eq.~(\ref{eq:bnsa2}), 
we can work out $B_n$, which becomes~:
\begin{eqnarray}
B_n \,&=&\, \frac{2 \, m_l}{Q} \, 
\sqrt{2 \, \varepsilon \, (1-\varepsilon )} \, 
\left( G_M^2 \,+\, \frac{\varepsilon}{\tau} \, G_E^2 \right)^{-1} \nonumber\\
&\times& \left\{ G_M  \left[ \frac{\sqrt{1 + \varepsilon} \sqrt{\tau}}
{\sqrt{1 + \varepsilon} \sqrt{1 + \tau} + \sqrt{1 - \varepsilon} \sqrt{\tau} } 
\,{\mathcal I} \left( A \right) - 
\sqrt{\frac{1 + \varepsilon}{2 \, \varepsilon}} 
\sqrt{\frac{\tau}{1 + \tau}} \, 
{\mathcal I} \left( A' \right)\right] \right. 
\nonumber \\
&+& \left. \frac{1}{\tau} G_E  
\left[ \frac{\sqrt{1 + \varepsilon} \sqrt{\tau} 
+ \sqrt{1 - \varepsilon} \sqrt{1 + \tau}}
{\sqrt{1 + \varepsilon} \sqrt{1 + \tau} + \sqrt{1 - \varepsilon} \sqrt{\tau} } 
\,\sqrt{\frac{1 + \varepsilon}{2 \, \varepsilon}} \,
{\mathcal I} \left( B \right) - 
\sqrt{\frac{\tau}{1 + \tau}} \, 
{\mathcal I} \left( B' \right)\right] 
\right\}  \nonumber \\
&+& \, {\mathcal{O}}(e^4) . 
\label{eq:bnsa3} 
\end{eqnarray}
Note that the soft part of $\tilde f_1$ (Eq.~(\ref{eq:f1qimsoft})) 
only enters in the amplitudes $\tilde G_M$ and $\tilde F_2$ as 
shown in Ref.~\cite{YCC04}, and does not enter into $B_n$ 
which is IR finite. One sees from Eq.~(\ref{eq:bnsa3}) 
that $B_n$ contains two terms : 
a first one in which the magnetic form factor $G_M$ is 
multiplied by a ``magnetic GPD'' ($H^q + E^q$), 
and a second one in which the electric form factor $G_E$ 
is multiplied by an ``electric GPD'' ($H^q - \tau E^q$). 
Furthermore, one notices that $B_n$ does not depend upon the GPD $\tilde H$.
\newline
\indent
To estimate the two-photon exchange amplitudes according to 
Eqs.~(\ref{eq:F3handbag},\ref{eq:F4handbag},\ref{eq:F5handbag}), 
we need to specify a model for the GPDs. As $B_n$ does not depend 
upon $\tilde H^q$, we only need to specify the model for the GPDs 
$H^q$ and $E^q$. 
Following Ref.~\cite{Rad98}, we use an unfactorized (valence) model in 
$x$ and $t$ for the GPD $H$ as~:
\begin{eqnarray}
\label{eq:gpdh}
H^q(x, 0, q^2) \,=\, q_v(x) \; 
\exp \left(- \frac{(1 - x) \, Q^2}{4 \, x \, \sigma } \right) , 
\end{eqnarray}
where $q_v(x)$ is the valence quark distribution.
In the following estimates we take the unpolarized parton distributions 
at input scale $Q_0^2$ = 1 GeV$^2$ from the 
MRST2002 global NNLO fit~\cite{mrst02} as~:  
\begin{eqnarray}
u_v &=& 0.262 \, x^{-0.69} (1 - x)^{3.50} 
\left( 1 + 3.83 \, x^{0.5} + 37.65 \, x \right), \nonumber \\
d_v &=& 0.061 \, x^{-0.65} (1 - x)^{4.03} 
\left( 1 + 49.05 \, x^{0.5} + 8.65 \, x \right) . \nonumber 
\end{eqnarray}
For the GPD $E$, whose forward limit is unknown, we adopt a valence 
parametrization multiplied with $(1 - x)^2$ 
to be consistent with the $x \to 1$ limit~\cite{Yuan03}. This yields~:  
\begin{eqnarray}
\label{eq:gpde}
E^q(x, 0, q^2) = \frac{\kappa^q}{N^q} (1 - x)^2  q_v(x)  
\exp \left(- \frac{(1 - x) Q^2}{4 \, x \, \sigma } \right) ,
\end{eqnarray}
where the normalization factors $N^u, N^d$ are chosen in such a way 
that the first moments of $E^u$ and $E^d$ at $Q^2 = 0$ 
yield the anomalous magnetic moments 
$\kappa^u = 2 \kappa^p + \kappa^n = 1.673$ and 
$\kappa^d = \kappa^p + 2 \kappa^n = -2.033$ respectively.  
Furthermore, the parameter $\sigma$ in 
Eqs.~(\ref{eq:gpdh},\ref{eq:gpde}) 
can be related to the average transverse momentum 
of the quarks inside the nucleon as $\sigma = 5 \, < k_\perp^2 > $. Its 
value has been estimated in Ref.~\cite{Die99} as~: 
$\sigma \simeq 0.8$~GeV$^2$, which we will adopt in the 
following calculations.  

\section{Results and discussion}
\label{sec:results}

\begin{figure}[h]
\epsfysize=11.5cm
\centerline{\epsffile{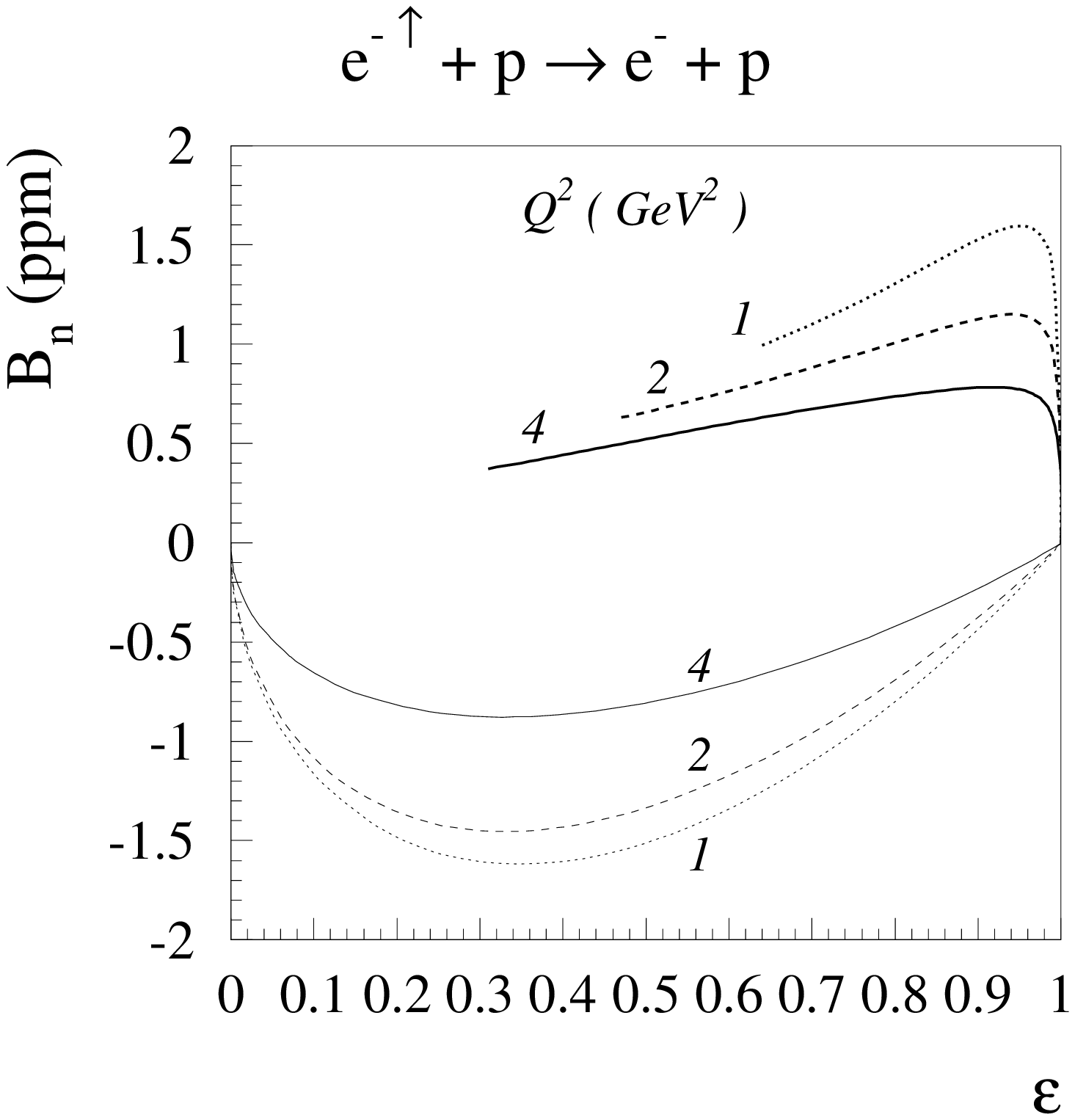}}
\caption{Beam normal spin asymmetry 
for elastic $e^- p$ scattering as function of $\varepsilon$  
at different values of $Q^2$ as indicated on the figure.  
The upper thick curves ($B_n > 0$) 
are the GPD calculations for the kinematical range where $s, -u > M^2$.
For comparison, the nucleon pole contribution is also displayed : lower 
thin curves ($B_n < 0$).}
\label{fig:bnsa_epsp}
\end{figure}
In Fig.~\ref{fig:bnsa_epsp}, we show our results for $B_n$ 
at several values of $Q^2$ (for $Q^2 > 1$~GeV$^2$) 
as a function of $\varepsilon$. 
One sees that the handbag calculation 
for the inelastic part to $B_n$ 
yields asymmetries which are forward peaked and are in the range 
of +1 ppm to +1.5 ppm. 
For comparison, the nucleon pole contribution is also displayed.  
For the proton form factors, 
we use the $G_{E p} / G_{M p}$ ratio as extracted from the 
polarization transfer experiments~\cite{Gayou02}. 
For $G_{M p}$, we adopt the parametrization of Ref.~\cite{Bra02}. 
One sees from Fig.~\ref{fig:bnsa_epsp} that the nucleon pole 
contribution to $B_n$ has a sign opposite to 
the inelastic one and also has a different energy 
dependence. Whereas the inelastic part peaks at large $\varepsilon$ (forward 
direction), the nucleon pole contribution is small in the forward region and 
reaches its maximum at $\varepsilon$ values around 0.3. 
As this nucleon pole contribution is well known, 
one can always add it to the data 
to extract the inelastic part from experiment (in analogy with what is 
usually done to extract the inelastic part of moments of nucleon 
structure functions). 
\begin{figure}[h]
\epsfysize=11.5cm
\centerline{\epsffile{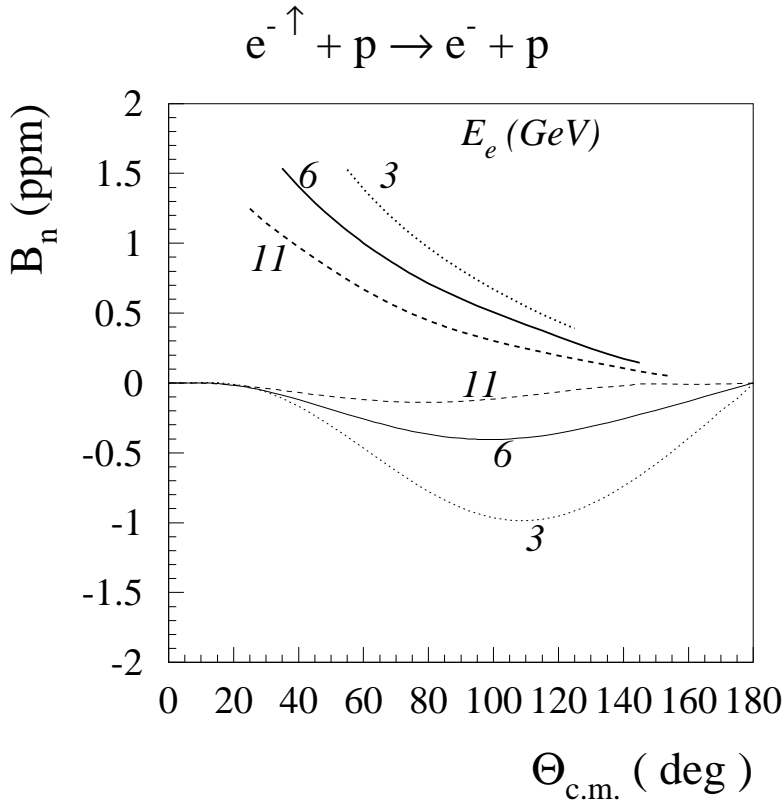}}
\caption{Beam normal spin asymmetry 
for elastic $e^- p$ scattering as function of the {\it c.m.} 
electron scattering angle at different electron beam energies 
as indicated on the figure.  
The upper thick curves ($B_n > 0$) 
are the GPD calculations for the kinematical range where $-u, Q^2 > M^2$. 
For comparison, the nucleon pole contribution is also displayed : lower 
thin curves ($B_n < 0$).}
\label{fig:bnsa_p}
\end{figure}
\newline
\indent
In Fig.~\ref{fig:bnsa_p}, we display our results for $B_n$ 
for elastic $e^{- \uparrow} p \to e^- p$ scattering 
at fixed beam energy as function of the elastic scattering {\it c.m.} angle 
in the energy range accessible at Jefferson Lab. 
One clearly sees that the forward angular range 
is a favorable region to get information on the inelastic part of $B_n$. 
As we have discussed before, this inelastic part is a direct measure of 
the imaginary part of the two-photon exchange amplitude. In the 
handbag calculation considered in this work, the real and imaginary parts 
are related through a dispersion relation. Hence a measurement of the 
inelastic contribution to $B_n$ would yield a useful cross-check on the 
handbag estimate for the real part, which was found crucial to resolve 
the discrepancy between Rosenbluth and polarization transfer 
experiments on the proton~\cite{YCC04}.
\begin{figure}[h]
\epsfysize=11.5cm
\centerline{\epsffile{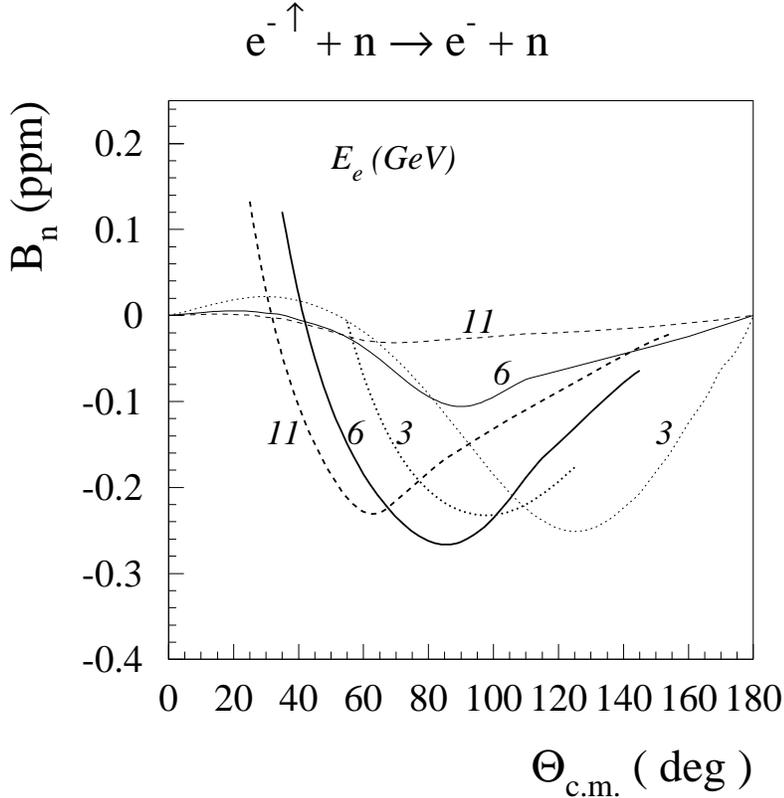}}
\caption{Beam normal spin asymmetry 
for elastic $e^- n$ scattering as function of the {\it c.m.} 
electron scattering angle at different electron beam energies 
as indicated on the figure.  
The thick curves are the GPD calculations 
for the kinematical range where $-u, Q^2 > M^2$. 
For comparison, the nucleon pole contribution is also displayed 
by the thin curves.}
\label{fig:bnsa_n}
\end{figure}
\newline
\indent
Within the handbag model described above, 
we also get a prediction for $B_n$ on the neutron. 
The two-photon exchange amplitude on the neutron is obtained 
by interchanging up and down-quark distributions in the 
corresponding expression for the proton. 
We show the neutron result for $B_n$ in Fig.~\ref{fig:bnsa_n}.
One sees that the inelastic part of $B_n$ for 
the neutron is much smaller than its proton counterpart. One can easily 
understand this from the expression of Eq.~(\ref{eq:bnsa2}) for $B_n$, 
which involves a term proportional to $G_M$ and a term proportional to 
$G_E$. For the proton case, both terms add with the same sign, whereas 
for the neutron case, $G_M$ has the opposite sign. 
This results in a partial cancellation in $B_n$. 
Furthermore, one sees from Fig.~\ref{fig:bnsa_n} that $B_n$ changes sign 
when going to the backward region 
(where the term proportional to $G_M$ dominates).  
In Fig.~\ref{fig:bnsa_n}, we also show the 
nucleon pole contribution to $B_n$ 
for the $e^{- \uparrow} n \to e^- n$ process. 
For the neutron form factors, 
we use the recent $G_{E n}$ parametrization of Ref.~\cite{War04}, 
and the $G_{M n}$ parametrization of Ref.~\cite{Kub02}. 
It is seen that also the nucleon pole contribution to $B_n$ for the neutron is 
suppressed, and is largely negligible compared to the inelastic one 
in the forward angular range.

\section{Conclusions}
\label{sec:concl}

In summary, we have developed in this work the formalism for 
elastic lepton-nucleon scattering with a lepton helicity flip, 
beyond the one-photon exchange approximation. We have shown 
that the imaginary part of the two-photon exchange amplitude 
can be accessed through the 
single spin asymmetry for a lepton beam polarized normal to the 
scattering plane. We provided an estimate for this normal beam spin 
asymmetry $B_n$ in a partonic framework, used before to  
evaluate the lepton helicity conserving amplitudes. 
In our calculation, the normal beam spin asymmetry at large momentum 
transfer is estimated through the scattering off a parton, 
which is embedded in the nucleon through a generalized parton distribution. 
Using phenomenological parametrizations for the GPDs, we  
found that for the proton, $B_n$ yields values around +1 ppm to +1.5 ppm 
in the few GeV beam energy range. 
In particular, we found that the forward angular  
range for $e^{- \uparrow} p \to e^- p$ scattering 
is a favorable region to get information on the inelastic part of $B_n$. 
Because in the handbag calculation considered here, real and imaginary 
parts are linked, a direct measurement of $B_n$ 
may yield a valuable cross-check of the estimate for the 
real part, which was found crucial in understanding the unpolarized 
cross section data for $e^- p \to e^- p$ at large momentum transfer.  
A measurement of $B_n$ can be performed   
by the same experiments that are set up to measure parity violation in 
$\vec e^{\, -} p \to e^- p$ scattering by choosing a normal 
polarization for the electron beam instead. 
This may open up a new experimental front to access 
the two-photon exchange amplitudes in elastic electron-nucleon scattering.

\section*{Acknowledgments}
This work was supported 
by the Italian MIUR and INFN, 
by the French Commissariat \`a l'Energie Atomique (CEA),
and by the U.S. Department of Energy under contracts 
DE-FG02-04ER41302 and DE-AC05-84ER40150. 

\appendix
\section{Relations between helicity amplitudes and invariant amplitudes 
for elastic electron-nucleon scattering}
\label{eq:app1}

In this appendix we express the amplitudes $\tilde F_3, \tilde F_4$ and 
$\tilde F_5$ which enter into the beam normal spin asymmetry of 
Eq.~(\ref{eq:bnsa2}) in terms of the helicity amplitudes for elastic 
lepton-nucleon scattering.  
For convenience, we introduce for the six independent helicity amplitudes 
of Eqs.~(\ref{eq:non-flip}, \ref{eq:flip}) 
(in the lepton-nucleon {\it c.m.} system) the shorthand notation~:
\begin{eqnarray}
T_1 \,&\equiv& \, T_{h'= +\frac{1}{2} \, \lambda_N' = +\frac{1}{2} \, ,\, 
h = +\frac{1}{2} \, \lambda_N = +\frac{1}{2}} , 
\hspace{1cm}
T_4 \equiv T_{h'=-\frac{1}{2} \,\lambda_N' =\frac{1}{2} \, , \, 
h=\frac{1}{2} \,\lambda_N =\frac{1}{2}} ,
\nonumber \\
T_2 \, &\equiv& \, T_{h'= +\frac{1}{2} \, \lambda_N' = -\frac{1}{2} \, ,\, 
h = +\frac{1}{2} \, \lambda_N = +\frac{1}{2}} , 
\hspace{1cm}
T_5 \equiv T_{h'=-\frac{1}{2} \,\lambda_N' =-\frac{1}{2} \, , \, 
h=\frac{1}{2} \,\lambda_N =\frac{1}{2}} ,
\nonumber \\
T_3 \, &\equiv& \, T_{h'= +\frac{1}{2} \, \lambda_N' = -\frac{1}{2} \, ,\, 
h = +\frac{1}{2} \, \lambda_N = -\frac{1}{2}} , 
\hspace{1cm}
T_6 \equiv T_{h'=-\frac{1}{2} \,\lambda_N' =\frac{1}{2} \, , \, 
h=\frac{1}{2} \,\lambda_N =-\frac{1}{2}} .
\label{eq:helampl}
\end{eqnarray}
The amplitudes $\tilde F_3, \tilde F_4$ and 
$\tilde F_5$ can be expressed in terms of the {\it c.m.} helicity amplitudes 
of Eq.~(\ref{eq:helampl}) as~:
\begin{eqnarray}
e^2 \, \frac{\tilde F_3}{M^2} \,&=&\,  \frac{1}{(s - M^2)} \,
\left\{ \, - T_1 
\,+\, \frac{2 M \, \sqrt{Q^2}}{\sqrt{M^4 - s u}} \, T_2 
\right. \nonumber \\
&& \left. \hspace{2.5cm}
+\, \left( \frac{(s^2 - M^4) - M^2 (s - u) }{(M^4 - s u)} \right) T_3
\, \right\} ,
\label{eq:f3thel} \\
e^2 \frac{\tilde{F}_4}{M}&=&\frac{M}{(s - M^2)}
\left[
- T_1\,+\,\frac{(s + M^2)}{M} \frac{\sqrt{Q^2}}{\sqrt{M^4 - s u}} \, T_2 
\right. \nonumber \\
&& \left. \hspace{2.5cm} +\,
\left( \frac{(s^2 - M^4) - M^2 (s - u) }{(M^4 - s u)} \right) T_3
\right] \nonumber \\
&+& \frac{\sqrt{Q^2}}{\sqrt{M^4 - s u}} \, \frac{M}{m_l} \, T_4 
\;+\;\frac{1}{2m_l}(T_5 - T_6),  
\label{eq:f4thel} \\
e^2 \frac{\tilde{F}_5}{M^2} &=&-\frac{2 \, M^2}{(s - M^2)^2}
\left[
-T_1\,+\,\frac{(s + M^2)}{M} \frac{\sqrt{Q^2}}{\sqrt{M^4 - s u}} \, T_2
\right. \nonumber \\
&& \left. \hspace{2.5cm}+\,\left(
\frac{s (s^2 + s u - 2 M^4) - M^4 (s - u)}{2 M^2 (M^4 - s u)}\right) T_3
\right] \nonumber \\
&-& \frac{(s + M^2)}{(s - M^2)} \, \frac{\sqrt{Q^2}}{\sqrt{M^4 - s u}} 
\, \frac{1}{m_l} \, T_4 
\;-\; \frac{1}{(s - M^2)} \, \frac{M}{m_l} \, (T_5 - T_6), 
\label{eq:f5thel}
\end{eqnarray}
where we keep only the leading term when taking $m_l \to 0$.

\end{document}